\documentclass[%
 aip,
floatfix,
 amsmath,amssymb,
 reprint,%
]{revtex4-1}

\usepackage{graphicx}
\usepackage{dcolumn}
\usepackage{bm}

\usepackage[utf8]{inputenc}
\usepackage[T1]{fontenc}
\usepackage{mathptmx}

\usepackage[version=4]{mhchem} 
\usepackage{hyperref}

\hypersetup{bookmarksnumbered=true,
  colorlinks = true, 	
  allcolors = blue,		
  breaklinks = true,	
  backref = true		
}
\begin{document}

\preprint{AIP/123-QED}

\title{Few-monolayer yttria-doped zirconia films: Segregation and phase stabilization
\vspace{10pt}}

\author{Peter Lackner}%
\affiliation{Institute of Applied Physics, TU Wien, 1040 Vienna, Austria}%

\author{Amy J. Brandt}
\affiliation{Institute of Applied Physics, TU Wien, 1040 Vienna, Austria}%
\affiliation{Department of Chemistry and Biochemistry, University of South Carolina, Columbia, SC 29208, USA}

\author{Ulrike Diebold}%
 
\author{Michael Schmid}%
\email{schmid@iap.tuwien.ac.at}
\affiliation{Institute of Applied Physics, TU Wien, 1040 Vienna, Austria}%

\date{\today}

\begin{abstract}
For most applications, zirconia (\ce{ZrO2}) is doped with yttria. Doping leads to the stabilization of the tetragonal or cubic phase, and increased oxygen ion conductivity. Most previous surface studies of yttria-doped zirconia were plagued by impurities, however. We have studied doping of pure, 5-monolayer \ce{ZrO2} films on Rh(111) by x-ray photoelectron spectroscopy (XPS), scanning tunneling microscopy (STM), and low-energy electron diffraction (LEED). STM and LEED show that the tetragonal phase is stabilized by unexpectedly low dopant concentrations, 0.5\,mol\% \ce{Y2O3}, even when the films are essentially fully oxidized (as evidenced by XPS core level shifts).
XPS also shows Y segregation to the surface with an estimated segregation enthalpy of $-23 \pm 4$\,kJ/mol.

\end{abstract}

\maketitle

\section{\label{sec:intro}Introduction}

Zirconia (\ce{ZrO2}) is an oxide with high thermal stability and favorable properties for many applications. \ce{ZrO2} is usually doped with yttria, which enhances the mechanical strength (fracture toughness).\cite{garvie_ceramicSteel,hannink_toughening} This is related to the control of the crystallographic modification by yttria doping and forms the basis for applications as an engineering material. By introducing oxygen vacancies into the lattice, yttria doping also increases the oxygen ion conductivity,\cite{minh_SOFCreview} which makes yttria-doped \ce{ZrO2} a solid-state electrolyte with high ionic (though very low electronic) conductivity at high temperatures, providing the basis for its use in gas sensors and solid oxide fuel cells (SOFCs).\cite{goodenough_electrolytes_2003} Due to the small size of Zr cations, oxygen-oxygen repulsion governs the stability of zirconia structures; with decreasing temperature, pure zirconia undergoes transformation from a cubic structure via tetragonal ($T < 2377\,^\circ$C) to monoclinic ($T < 1205\,^\circ$C), in order to maximize O--O distances while maintaining short Zr--O bonds. \cite{kisi_crystal_1998} Already in the beginning of the 20\textsuperscript{th} century, it was found \cite{ruff_1929} that non-monoclinic phases can be stabilized by doping, e.g., with magnesia, thoria, or -- most commonly -- yttria (YSZ, yttria-stabilized zirconia). Added trivalent Y$^{3+}$ replaces tetravalent Zr$^{4+}$; for charge compensation, one oxygen vacancy is created per two Y atoms. \cite{goetsch_zirconia_2016,kisi_crystal_1998} These vacancies form the basis for the high oxygen ion conductivity at high temperatures. Distortions around the vacancy and an overall lattice expansion decrease the average O--O repulsion, stabilizing the phases otherwise present at high temperatures only. Below 1.5\,mol\% \ce{Y2O3},
\footnote{For the relation between mol\% \ce{Y2O3} and Y concentration or Y/Zr ratio in the film, see Ref.\ \onlinecite{goetsch_zirconia_2016}}
 zirconia remains monoclinic.
Concentrations above 7.5\,mol\% \ce{Y2O3} stabilize the cubic phase.\cite{scott_phase_1975} Between 1.5\,mol\% and 7.5\,mol\%, a more complicated behavior is reported. Depending on the preparation parameters, either a mixture of the cubic and monoclinic phases, tetragonal and cubic phases, or only the tetragonal phase is formed. \cite{scott_phase_1975,chen_thermodynamic_2004} Tetragonal \ce{ZrO2} can accommodate strain by changing the crystal phase and/or orientation, which leads to the above mentioned high mechanical strength (tetragonal zirconia polycrystals TZP, zirconia toughened ceramics ZTC). \cite{garvie_ceramicSteel,hannink_toughening} In nanoparticles, the tetragonal and the cubic phase can form at lower dopant concentrations or even without doping. This stabilization is mainly due to the introduction of oxygen vacancies.\cite{shukla_mechanisms_2005,lacknerPCCP}

Upon annealing YSZ at high temperatures in ultrahigh vacuum (UHV), \cite{vonk_SXRD} or \ce{O2}, \cite{theunissen_segregation_1989,de_ridder_LEIS} yttrium can diffuse and segregate to the surface and grain boundaries. \cite{nowotny_chargeTransfer3} This can lead to local phase transformations, as the yttrium content in some regions increases while it decreases in others. \cite{witz_XRD, chao_enhanced_2013} Segregation is a well-studied topic for YSZ, as both, impurity (mainly Si, but also Ca and Na) and Y dopant segregation can influence material properties such as the oxygen exhange \cite{de_ridder_LEIS_O2} or the selectivity in chemical reactions, as shown for the example of formate oxidation. \cite{lahiri_STM_XPS} Many studies agree that surface segregation in YSZ is dominated by impurities. \cite{hughes_XPS,de_ridder_LEIS,chaim_silicateGlass,bernasik_XPS} The surface region then usually consists of silicates, with the subsurface region enriched in yttria; it seems that the interface stabilizes the yttria below. \cite{de_ridder_LEIS,hughes_XPS} Only few experimental results are available for yttria segregation with little influence from impurities, \cite{lahiri_STM_XPS,vonk_SXRD} yet yttria surface segregation is also found in these. A study on YSZ single crystals with ALD-deposited surface layers of increased Y content showed that oxygen incorporation, a typical rate-limiting step for SOFCs, is increased with increasing yttrium concentration near the surface, \cite{chao_enhanced_2013} while a silicon-containing surface layer leads to a decrease. As studies of Y segregation on clean surfaces are rare, neither the concentration profile nor the impact on applications are completely understood. Density functional theory (DFT) studies do not agree whether Y enrichment should occur in the uppermost layer, \cite{mayernick_dft_2010} or in the layer immediately below \cite{wang_yttrium_2008}; a recent surface x-ray diffraction study \cite{vonk_SXRD} suggests surface enrichment.

For a controlled atomistic study of the surface region of YSZ, single crystals can be used. These are typically cubic with a doping level of 8--10\,mol\% \ce{Y2O3}. As YSZ is an electronic insulator with a wide band gap, \cite{gotsch_bandgap} surface science studies at room temperature (RT) are difficult. Morrow et al. used high-temperature scanning tunneling microscopy (STM) to study the surface at 300\,$^\circ$C; the existence of a tunneling current at this temperature was attributed to both, electronic and ionic currents. \cite{morrow_STM} At RT, the insulating nature of the material can be circumvented by measuring on thin films. In the present work, we build on our previous work on pure zirconia thin films grown on Rh(111) single crystals; these films have been thoroughly characterized. \cite{lackner_surfacestructure, lacknerPCCP, lacknerSMSI} We showed that films with a thickness of five monolayers (ML) or more form bulk-like structures -- either the tetragonal or the monoclinic phase -- depending on the annealing temperature. \cite{lackner_surfacestructure} \ce{ZrO2} films annealed at temperatures below 730\,$^\circ$C are tetragonal. When annealing the films at 850\,$^\circ$C, holes down to the Rh substrate appear in the films, i.e., the zirconia starts dewetting the substrate. Simultaneously, the films transform to the monoclinic phase. Low energy electron diffraction (LEED) and STM can be used to quickly characterize the film structure, as tetragonal films exhibit a $(2 \times 1)$ periodicity with respect to cubic \ce{ZrO2}(111), while monoclinic films show a distorted $(2 \times 2)$ structure (angles differ from 60 and 120$^\circ$). In STM images, the tetragonal structure exhibits rows with a distance of 0.63\,nm; the Zr--Zr distance of 0.36\,nm within the rows is more difficult to resolve.\cite{lackner_surfacestructure}

X-ray photoelectron spectroscopy (XPS) showed that tetragonal films are slightly reduced (off-stoichiometry $< 2\%$). \cite{lacknerPCCP} Oxygen vacancies stabilize the tetragonal phase and are positively charged w.r.t.\ the unperturbed lattice (V$_\mathrm{O}^{\bullet\bullet}$ in Kröger-Vink notation). This charge shifts the electrostatic potential, resulting in substantial binding energy ($E_\mathrm{B}$) differences ($\Delta E_\mathrm{B} \ge 1.3$\,eV) compared to monoclinic zirconia films, which are (close to) fully oxidized. Zr 3d$_{5/2}$ levels of reduced, tetragonal zirconia were found between 183.4 and 182.6 eV, and for the monoclinic film between 181.8 and 181.6 eV. It should be noted that this shift, induced by different alignment of the oxide bands with respect to the Fermi level, is opposite to the usual chemical shift caused by different oxidation states; it is of purely electrostatic origin. The Zr in the films remains in the 4+ charge state. \cite{lacknerSMSI, lacknerPCCP} When tetragonal films start dewetting the Rh substrate, the Rh substrate becomes exposed and can act as a catalyst for \ce{O2} dissociation;\cite{lacknerPCCP} the activated oxygen oxidizes the tetragonal zirconia film, and it transforms to the thermodynamically stable phase of stoichiometric \ce{ZrO2}, i.e., monoclinic zirconia.

In section \ref{ssec:tetra} of this work we show how to create yttria-doped 5 ML-thick zirconia films by Y deposition on \ce{ZrO2} and annealing in oxygen. The stability of tetragonal zirconia thin films is substantially increased by Y incorporation; already 0.5\,mol\% \ce{Y2O3} is sufficient to prevent formation of the monoclinic phase. Our approach provides a well-defined model system for yttria-doped \ce{ZrO2}, without impurities and with a well-defined Y content. In section \ref{ssec:xps}, the electronic structure of the films is studied with XPS and compared to undoped zirconia films. In section \ref{ssec:segregation}, XPS results are used to extract the Y segregation behavior.

\section{\label{sec:expt}Experimental Methods}

We used a two-chamber UHV system with STM, LEED, and XPS capabilities in the analysis chamber ($p_\mathrm{base} < 6 \times 10^{-11}$\,mbar). Sample preparation was conducted in the preparation chamber ($p_\mathrm{base} < 10^{-10}$\,mbar), which is connected to the analysis chamber via a gate valve. The preparation chamber contains an electron-beam evaporator for Y (Omicron EFM 3), a UHV-compatible Zr sputter source, \cite{lacknerSputterSource} an Ar$^+$ sputter gun, and an e-beam heating stage for the sample. The setup is described in more detail in Ref.\ \onlinecite{lackner_surfacestructure}; details about the XPS measurement setup in this UHV system as well as the data fitting procedure using the program CasaXPS are found in Ref.\ \onlinecite{lacknerPCCP}. LEED images were processed with dark frame and flat field correction\cite{koller_2002} to suppress artifacts such as the grid structure of the LEED optics.

5 ML-thick \ce{ZrO2} thin films were sputter-deposited ($p_\mathrm{Ar} \approx 6.5 \times 10^{-6}$\,mbar, $p_\mathrm{O2} = 1.5 \times 10^{-6}$\,mbar) at room temperature on a Rh(111) single crystal substrate (diameter 9 mm, thickness 2 mm; from MaTecK, Germany). The films were then post-annealed at 720\,$^\circ$C in \ce{O2} to form an ordered, tetragonal structure. \cite{lackner_surfacestructure} Unless noted otherwise, annealing was always done in \ce{O2} at $p_\mathrm{O2} = 5 \times 10^{-7}$\,mbar for 10 min. To dope these films with Y, either 0.05 ML or 0.2 ML (corresponding to 0.5 and 2\,mol\% \ce{Y2O3} with respect to films with 5 ML thickness) were deposited from an Y wire (1\,mm diameter, 99.9\% purity, handled in protective Ar atmosphere to avoid excessive oxidation or ignition). We define a deposited monolayer (ML) as one atom per Zr atom in the surface layer ($8.9 \times 10^{14}$\,cm$^{-2}$). Yttrium was deposited in an \ce{O2} background ($p_\mathrm{O2} = 5 \times 10^{-7}$\,mbar) at various temperatures, see below. As the sample could not be heated at the position in front of the evaporator, the samples were preheated; then, Y was deposited while cooling. The deposition rate of Y was measured by a water-cooled, retractable quartz crystal microbalance moved to the sample position before the actual deposition. Additionally, we also verified the Y deposition rate in an experiment where we evaporated a defined amount of Y on Rh(111) and measured the coverage with STM. Sample temperatures were measured with a thermocouple attached to the sample holder and calibrated at T > 800\,$^\circ$C using a disappearing-filament pyrometer. At lower $T$, the temperature differences between the sample and sample holder are extrapolated. We estimate the temperatures obtained by this procedure to be accurate within $\pm 30\,^\circ$C.

\section{\label{sec:results}Results}

\subsection{\label{ssec:tetra}Structure: Stabilization of the Tetragonal Phase}

\begin{figure*}
\includegraphics[width=15cm]{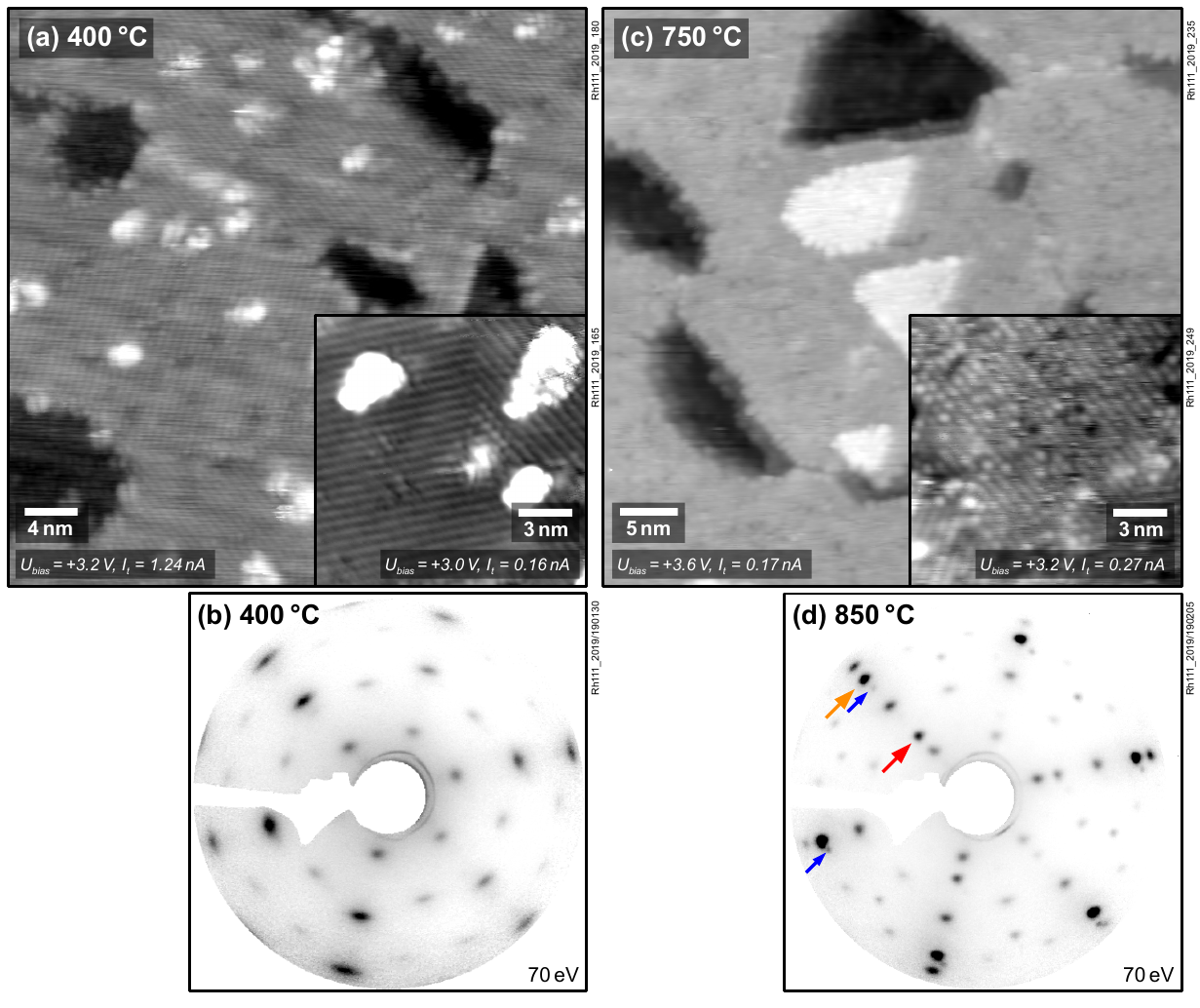}%
\caption{\label{fig:tetra}Preparation of 5 ML-thick yttria-doped \ce{ZrO2} films. (a) STM image of clusters after the deposition of 0.05 ML of Y on tetragonal zirconia. The small-area image (inset) shows that the typical row structure of tetragonal \ce{ZrO2} is unaffected by the deposition. (b) The LEED image of this surface confirms that the tetragonal phase remains stable after deposition. (c) STM image after annealing the film in (a,b) at 750\,$^\circ$C. The film is still tetragonal and not yet broken, and larger islands appear at the surface. The row structure is still present, and a large number of point defects appear (inset). (d) After annealing at 850\,$^\circ$C, Rh spots (orange arrow) indicate that the film dewets the substrate, while the tetragonal structure is unchanged. (Blue arrows mark weak spots originating from a different Rh crystallite near the edge, and the red arrow points at a spot of the $(2\times 1)$ O/Rh superstructure.\cite{kohler_high-coverage_2004})}
\end{figure*}

The standard preparation of tetragonal yttria-doped films started with an ordered, closed, 5 ML-thick zirconia film in the tetragonal phase (annealed at $T = 720\,^\circ$C). After confirming the successful preparation of the structure with LEED and STM, sub-monolayers of yttrium were deposited on top of the film at $T = 400\,^\circ$C in \ce{O2}. STM measurements revealed cluster formation at this temperature; Figure \ref{fig:tetra}a shows the surface directly after deposition of 0.05 ML Y. The inset of Figure \ref{fig:tetra}a as well as the LEED pattern in Fig. \ref{fig:tetra}b confirmed the unchanged tetragonal structure of the film after deposition (cf.\ Ref.\ \onlinecite{lackner_surfacestructure}). The height of the clusters ($\approx 0.3$\,nm) is comparable to the interlayer distance of cubic \ce{ZrO2}(111) or \ce{Y2O3}(111). We could not resolve the atomic structure of the clusters, probably because of their small size (flat areas $<2$\,nm). The clusters remained stable when annealed at 550\,$^\circ$C. However, after annealing 0.05 ML of Y on tetragonal zirconia at $T = 750\,^\circ$C (Fig. \ref{fig:tetra}c), larger islands formed at the surface. In STM, their surface appeared like that of the \ce{ZrO2} film, with rows characteristic for tetragonal \ce{ZrO2}. The triangular and hexagonal shapes of the islands are also found for undoped zirconia films. \cite{lackner_surfacestructure} We therefore conclude that the material of the clusters had diffused into the film, resulting in zirconia (probably doped) being expelled to the surface, forming a partial 6\textsuperscript{th} layer. In principle, part of the island material might also have spilled out from holes in the film (which do not reach the substrate at this temperature). Since the area fraction of the islands (5\%) is compatible with the amount of deposited yttria, it is unlikely that the material in the islands originates from the holes. Also, the islands are not preferentially located close to these holes, as typically observed for material flowing out from holes. \cite{lacknerSMSI} Diffusion of Y into the film was confirmed by XPS, see below.

When annealing at even higher temperatures ($T = 850\,^\circ$C; Figure \ref{fig:tetra}d), holes in the film revealed the Rh(111) substrate below. This can be inferred from the appearance of bright Rh spots in LEED (marked with orange arrows in Figure \ref{fig:tetra}d) as well as STM images (not shown). For a pure zirconia film, access to Rh would lead to full oxidation of the film and thus to a transformation to the monoclinic phase. \cite{lacknerPCCP} However, with added Y, dewetting and exposing the Rh substrate was not accompanied by such a phase transformation; LEED still showed the $(2 \times 1)$ pattern of tetragonal \ce{ZrO2}, not the more complex pattern \cite{lackner_surfacestructure} of the monoclinic phase. Considering the low amount of Y deposited, this result came somewhat unexpected. Assuming a homogeneous distribution of Y across all five layers of the film, the material is doped with only 1 at\% of Y or 0.5\,mol\% \ce{Y2O3}. This is quite low compared with $>1.5$\,mol\% needed for stabilization of the tetragonal phase in the bulk. As expected, also films doped with a higher amount of Y, 0.2 ML (2\,mol\% when assuming a homogeneous distribution; deposited at 550\,$^\circ$C in \ce{O2}) remained tetragonal up to the highest annealing temperature we tested (950\,$^\circ$C; not shown). 

Although the film remained tetragonal, the number of point defects at the surface increased after Y deposition and annealing, as becomes apparent by comparing the STM images in the insets of Figs.\ \ref{fig:tetra}a and b. Two types of point defects appeared: Bright species and a few dark holes. It is tempting to interpret the bright species as Y ions; nevertheless it is also possible that the bright features are due to an electronic effect caused by oxygen vacancies.

\begin{figure}
\includegraphics[width=7cm]{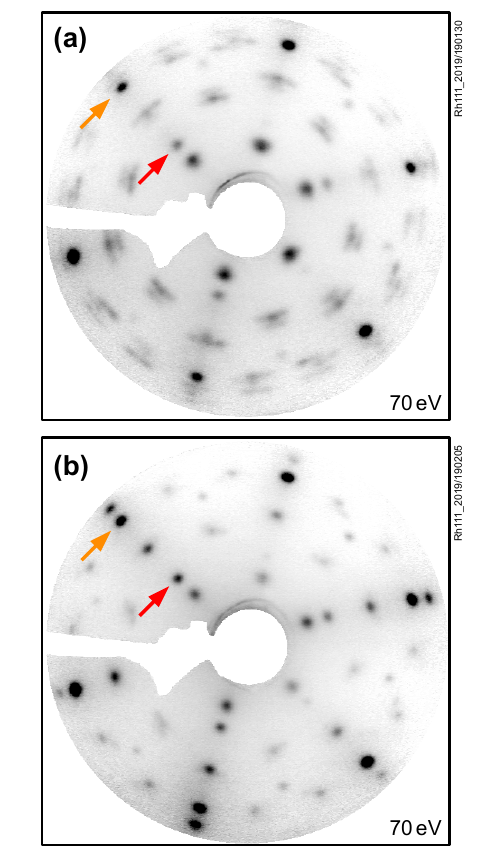}%
\caption{\label{fig:mono}LEED patterns of the transformation of a monoclinic film to the tetragonal phase. (a) Original monoclinic zirconia film, with typical elongated and multiple spots due to the unit cell angles deviating from 60$^\circ$. \cite{lackner_surfacestructure} The orange arrow marks a Rh spot, and the red arrows points at a spot of the $(2\times 1)$ O/Rh superstructure. (b) After deposition of 0.2 ML Y and annealing at 850\,$^\circ$C, the film transforms to the tetragonal phase. Some LEED spots are slightly smeared out, possibly due to remaining small, monoclinic areas.}
\end{figure}

In a different experiment, a monoclinic \ce{ZrO2} film was prepared by annealing at 850\,$^\circ$C. The nominal film thickness was 5 ML; due to the holes where the substrate was uncovered (dewetting), the local film thickness was 6--9 ML. On this monoclinic film, 0.2 ML of Y were deposited at 550\,$^\circ$C in \ce{O2}. The deposition did not change the structure of the film. However, post-annealing at 850\,$^\circ$C in $5 \times 10^{-7}$\,mbar \ce{O2} led to a complete transformation back to the tetragonal phase, see Figure \ref{fig:mono}. Without Y, such an additional annealing step would not change the structure of the film. A monoclinic $\rightarrow$ tetragonal transformation is also possible without Y-doping by annealing under highly reducing conditions (950\,$^\circ$C in UHV\cite{lackner_surfacestructure}), which introduce oxygen vacancies. With added yttrium, the phase transition also occurs under oxidizing conditions and at lower temperatures.

\subsection{\label{ssec:xps}Photoelectron Spectroscopy: Core level shifts}

\begin{figure*}
\includegraphics[width=17cm]{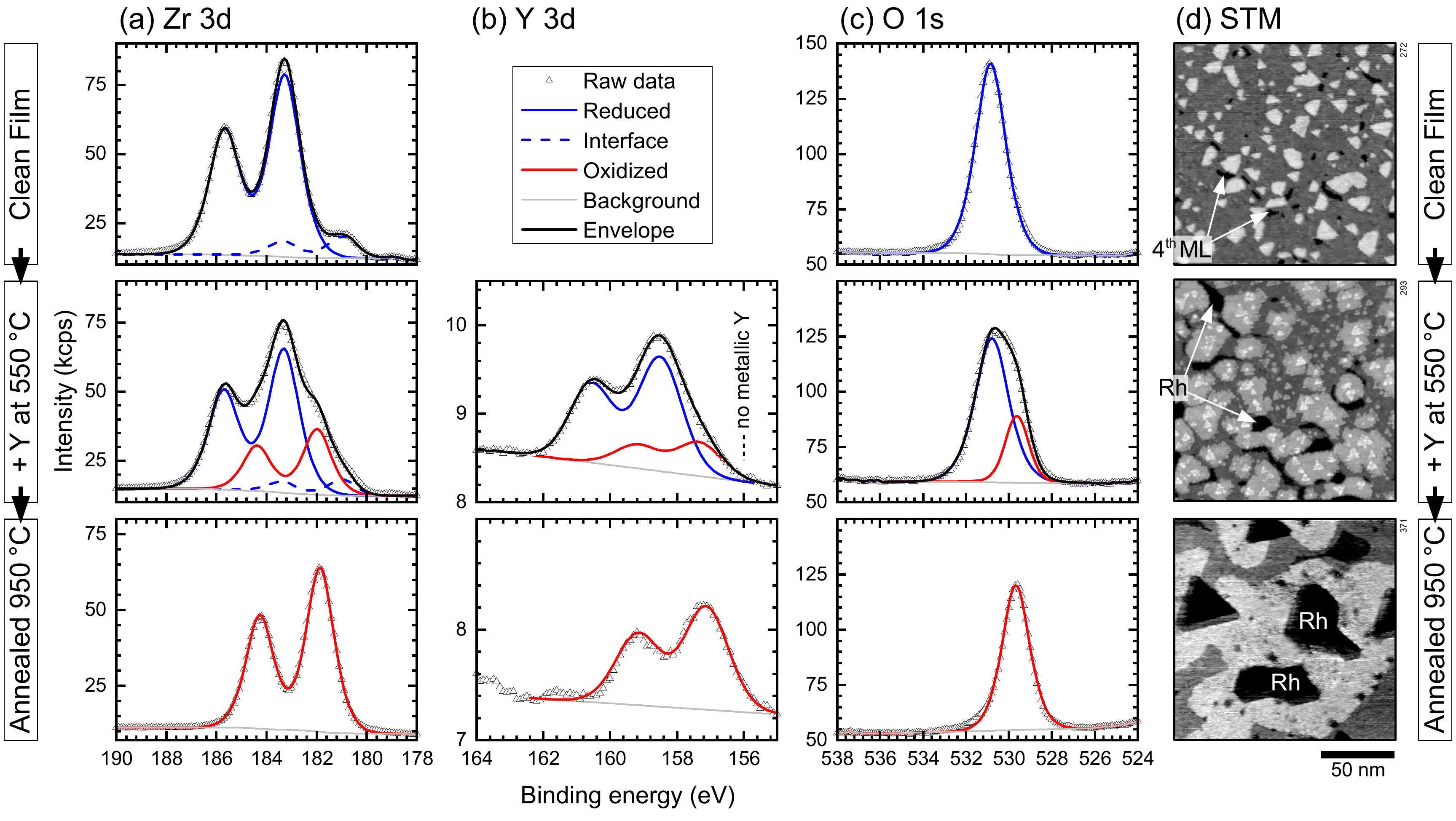}%
\caption{\label{fig:xps}XPS measurements and fits for 0.2 ML Y/\ce{ZrO2}, (a) Zr 3d, (b) Y 3d, and (c) O 1s region. (d) STM images for these preparations. The top panels are before Y deposition, the middle panels are after Y deposition at 550\,$^\circ$C, and the lower panels are after annealing at 950\,$^\circ$C in \ce{O2}.}
\end{figure*}

X-ray photoelectron spectroscopy (XPS) was measured for freshly prepared, tetragonal zirconia films, after deposition of Y, and after each of several annealing steps. Results from a film with 0.2 ML Y are shown in Figure \ref{fig:xps}. 

For the pure, tetragonal zirconia films used in this study, we found the Zr 3d$_{5/2}$ and O 1s levels at binding energies of $E_\mathrm{B}$ = 183.3 eV and 530.9 eV, respectively -- in the range observed previously for 5 ML-thick tetragonal zirconia films. \cite{lacknerPCCP} The binding energies are at the high side of this range, however, pointing towards a rather high oxygen vacancy concentration (in the order of $\approx 2$\%). \cite{lacknerPCCP} These binding energies were unaffected by Y deposition at T = 400\,$^\circ$C. Similar to pure zirconia films, annealing at higher temperatures led to holes in the films reaching down to the substrate (local dewetting). This resulted in an oxidation of the film by oxygen spillover from the metal, and thus the binding energy shifted to lower values. \cite{lacknerPCCP} The onset temperature of dewetting and oxidation of the film depends on the details of the preparation; for Y deposition at $T \approx 400\,^\circ$C, oxidation was encountered only after post-annealing at 750\,$^\circ$C. However, when Y was deposited at sufficiently high temperatures (T = 550\,$^\circ$C), as in the experiment shown in Figure \ref{fig:xps}, we found a few holes reaching down to the substrate (local dewetting) already after deposition. For this film, all XPS core levels (Zr, Y, and O) can be fitted by two contributions, one fixed to the binding energies of the reduced film (same Zr and O binding energies as before Y deposition), and one at a lower $E_\mathrm{B}$. This indicates partial oxidation of the film. Additional annealing at 750\,$^\circ$C in $5 \times 10^{-7}$\,mbar \ce{O2} converted the whole film into an essentially fully oxidized state (Zr 3d$_{5/2}$ at 181.9 eV).

For Y 3d$_{5/2}$, a typical $E_\mathrm{B}$ value of the main component (corresponding to the reduced film) of 158.5 eV was found after deposition. As expected for deposition in $5 \times 10^{-7}$\,mbar \ce{O2}, no metallic Y component was present. The core level shifts between the initially reduced films and the more oxidized films are similar for the Y 3d$_{5/2}$ and Zr 3d$_{5/2}$ levels: From the reduced to the oxidized film after annealing at 750\,$^\circ$C, the binding energies decreased by 1.2 and 1.3 eV, respectively. At 880\,$^\circ$C, Y 3d$_{5/2}$ shifted by an additional $-0.1$\,eV. Annealing at 950\,$^\circ$C led to another $E_\mathrm{B}$ shift of $-0.1$\,eV of both peaks. In total, both peaks shifted by $-1.4$\,eV. A film with lower doping (deposition of 0.05\,ML Y) showed approximately the same concerted shifts of Y and Zr, but the Y spectra were more noisy due to the low coverage, so the peak fitting is less accurate.  O 1s shifted by 0.2 eV less than Zr 3d$_{5/2}$ and Y 3d$_{5/2}$ in the 0.2 ML preparation, and 0.3 eV less in the 0.05 ML preparation, when comparing the preparations before Y deposition and after annealing at 750\,$^\circ$C (and above). None of these shifts were accompanied with changes of the crystallographic structure, which was always tetragonal, as confirmed by STM and LEED.

\subsection{\label{ssec:segregation}Y Segregation}

The intensity ratio of Y 3d and Zr 3d peaks can be used to track the diffusion of Y in the film. Using this ratio avoids strong effects from dewetting, which leads to increased film thickness for the remaining film and therefore lower intensities (due to the very similar kinetic energy of the photoelectrons, both, the Y and the Zr signals are equally affected at least in case of a homogeneous Y distribution). Since the Y/Zr ratio depends on the exact amount of Y deposited, in this section we only discuss {\em changes} of the Y/Zr intensity ratio before and after annealing; in contrast to the Y/Zr ratios themselves these changes are insensitive to inaccurate initial Y coverages.

After annealing a tetragonal zirconia film with 0.2 ML of Y at 750\,$^\circ$C, the Y 3d/Zr 3d intensity ratio dropped to 93\% of the original ratio, yet then remained nearly constant with 90\% after annealing at 880\,$^\circ$C and to 91\% after annealing at 950\,$^\circ$C. Due to the low intensity of the Y 3d doublet, the changes observed above 750\,$^\circ$C lie within the error bars. A film with 0.05 ML Y deposition showed a similar intensity ratio drop to 91\% after annealing at 880\,$^\circ$C. The drop in intensity could either indicate that yttria diffused from the clusters, which formed at the surface during Y deposition (see above), into the film, or it diffused over the surface to form larger, 3D yttria clusters, thick enough to attenuate some of the Y 3d signal. STM confirmed the first interpretation, as no 3D clusters were found at the surface after annealing at elevated temperatures. Thus, it can be assumed that Y is incorporated into the film. 

\begin{figure}[b]
\includegraphics[width=8.5cm]{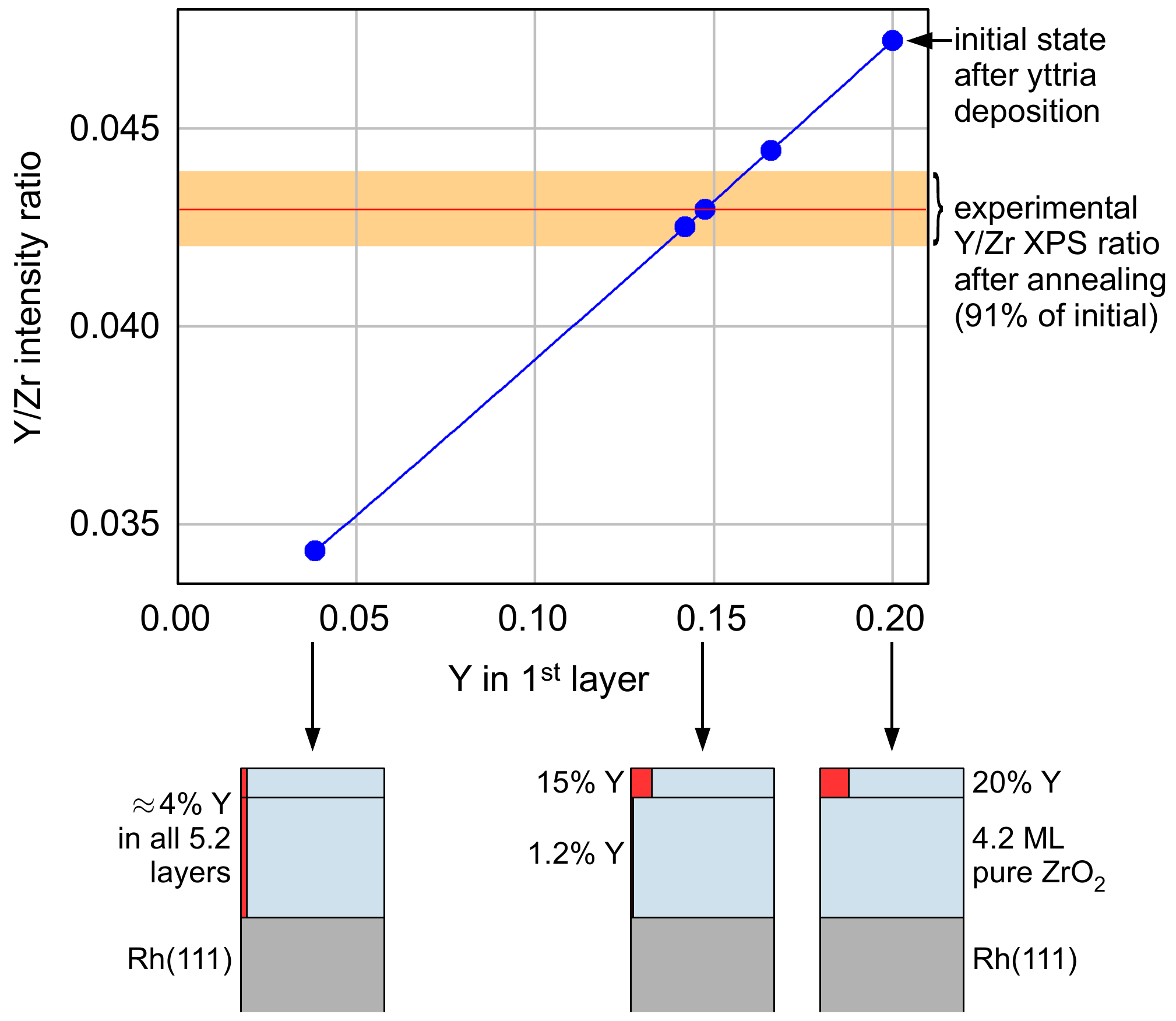}%
\caption{\label{fig:sessa}Simulated Y 3d/Zr 3d intensity ratios (blue) after yttria deposition (rightmost point), and with different degrees of Y enrichment in the surface layer. The total Y content is the same in all cases. The leftmost data point assumes homogeneous distribution of Y in the 5.2\,ML thick film. The decrease of the Y/Zr intensitiy ratio upon annealing as found experimentally is given by a horizontal, red line, with the orange area symbolizing the uncertainty. The corresponding concentration profiles (assuming that only the first layer can have a different Y concentration than the others) are shown at the bottom.%
}
\end{figure}

To estimate the distribution of Y in the film, we compare the measured intensity ratios Y 3d/Zr 3d to ratios simulated with the program SESSA. \cite{smekal_sessa} For these simulations, we assumed that the overall Y content of the film does not change upon annealing. For the oxygen concentrations in the layers, full oxidation was assumed, i.e. $x  $ZrO$_2 + y $YO$_{1.5}$, and the bandgap of the oxide layers was set to 5\,eV. We did not simulate the partially filled first layer as such but rather took a single layer (0.3\,nm thick) with appropriate average concentrations as the first layer (see bottom right of Fig.\ \ref{fig:sessa} for 0.2\,ML deposited on 5\,ML ZrO$_2$), since, in our experience, this leads to better accuracy.

When distributing 0.2 ML YO$_{1.5}$ homogeneously in all layers (resulting in a composition of Zr$_{0.96}$Y$_{0.04}$O$_{1.98}$, leftmost point in Fig.\ \ref{fig:sessa}), compared with the same amount of YO$_{1.5}$ in the first layer only, the Y/Zr intensity ratio drops to 73\%. Comparison with the experimental drop to $\approx 91$\% shows that, on average, Y does not diffuse deep into the film, but most Y stays near the surface of the film. Assuming Y enrichment in the first layer only and a constant concentration in the layers below, the experimental change of the Y/Zr intensity ratio can be reproduced with Zr$_{0.85}$Y$_{0.15}$O$_{1.93}$ in the first layer and  Zr$_{0.988}$Y$_{0.012}$O$_{1.994}$ below (see Fig.\ \ref{fig:sessa}). The experimental result is not compatible with Y enrichment in the second layer only, and the surface being pure \ce{ZrO2}, as suggested in a DFT study \cite{wang_yttrium_2008}: Moving all Y to the second layer would cause a decrease of the Y intensity to 85\%, a stronger decrease than observed experimentally, and the decrease would be even more when assuming some Y in the layers below.

To examine whether Y stays near the surface due to kinetic limitations, we have performed an experiment where the initial position of Y was below the surface: 2.4 ML of zirconia were deposited with the usual parameters, followed by 0.2 ML Y deposition at room temperature in \ce{O2}, and an additional 2.4 ML of zirconia. When annealing this film at 750\,$^\circ$C, the Y~3d/Zr~3d ratio increased by 24\%, and stayed constant after annealing at 880\,$^\circ$C and 950\,$^\circ$C. This again demonstrates Y segregation to the surface region, and the simulation can reproduce this intensity ratio for essentially the same surface concentration as above, Zr$_{0.86}$Y$_{0.14}$O$_{1.93}$ in the first layer and  Zr$_{0.984}$Y$_{0.016}$O$_{1.992}$ below. The good agreement with the concentration profile obtained after yttria deposition at the top and annealing indicates that Y diffusion is fast enough for equilibration. The high level of agreement also means that Y does not float up during room-temperature deposition of the upper 2.4 ML zirconia, but it gets buried by the zirconia (otherwise the increase of the Y/Zr intensity ratio would be less).

\section{\label{sec:disc}Discussion}

After deposition of yttrium on zirconia films at $5 \times 10^{-7}$\,mbar \ce{O2}, the small size of the clusters found by STM suggests the formation of small \ce{Y2O3} aggregates; we have no evidence for immediate diffusion of Y into the film at 400\,$^\circ$C. The enthalpy of formation of \ce{Y2O3} is $\approx -1920$\,kJ/mol. \cite{swamy_Y2O3_thermodynamics}  This corresponds to $-6.6$\,eV per O atom, or, taking this value as a chemical potential, an oxide-metal equilibrium at an oxygen partial pressure of $10^{-89}$\,bar at 400\,$^\circ$C. Therefore, metallic Y can be safely excluded (in agreement with XPS showing no metallic Y). After deposition or annealing at higher temperatures, larger, triangular or hexagonal islands appear, with the surface structure typical for tetragonal zirconia. \cite{lackner_surfacestructure} This is accompanied by a lowering of the XPS peak ratios of Y 3d/Zr 3d, which shows that Y is incorporated into the film. However, Y stays preferably in the surface region of the film, i.e., the topmost layer, even after annealing at temperatures of up to 950\,$^\circ$C. We cannot exclude some Y enrichment in the layer below, but the results do not support Y enrichment in the second layer only, as suggested in a DFT study.\cite{wang_yttrium_2008} When Y is placed in the middle layer of the film already during deposition, it also segregates to the surface region after annealing.   In contrast to previous experimental studies, our films contain no traces of impurities (down to the detection limit of XPS). \cite{lacknerSputterSource} Since it has been suggested that impurities like Si, Na or Ca enhance yttrium segregation, \cite{de_ridder_LEIS} we consider our study the first confirmation of Y surface segregation where the influence of impurities can be definitely ruled out.

We can use the Langmuir-McLean equation \cite{schmid_segregation_2002} to estimate the segregation enthalpy of the YO$_{1.5}$/\ce{ZrO2} system. Assuming that only the surface layer is enriched in Y, from the concentrations mentioned above we obtain $\Delta H_\mathrm{segr} = -23 \pm 4$\,kJ/mol.

Adding 0.5\,mol\% \ce{Y2O3} to 5 ML-thick zirconia films is sufficient to stabilize them in the tetragonal phase. It should be noted that this is an average concentration; due to Y segregation the surface concentration is substantially higher, while the subsurface layers contain less than 0.5\,mol\% \ce{Y2O3}. For comparison, in bulk zirconia, more than 1.5\,mol\% is needed for the stabilization of non-monoclinic phases. In Ref.\  \onlinecite{lacknerPCCP}, the amount of oxygen vacancies stabilizing the tetragonal phase in 5 ML-thick zirconia films was estimated to be $\lesssim 2$\%. To induce an oxygen vacancy concentration of 1\%, an Y concentration of 2\%, corresponding to 1\,mol\% \ce{Y2O3} would be required. There can be different reasons why the tetragonal structure is more stable in thin films than in the bulk: (i) It is possible that the surface energy of the tetragonal phase is lower, as suggested previously. \cite{garvie_stabilization} Calculated values of the surface energies for the relevant surface orientations \cite{christensen_carter} are very similar for the two phases, however. (ii) The interface to the substrate below may stabilize the tetragonal phase. (iii) Even the oxidized tetragonal yttria-doped films may still contain additional oxygen vacancies (beyond those introduced by Y doping), stabilizing the tetragonal phase. Currently, we cannot decide whether (ii) and/or (iii) are the main factors extending the stability of the tetragonal phase to lower Y concentrations.

As for pure \ce{ZrO2} films, \cite{lacknerPCCP} all XPS core level shifts depend on the oxygen vacancies present in the film. In the current case of yttria-doped films, we have to distinguish between oxygen vacancies caused by doping (keeping the Zr$_{x}$Y$_{y}$O$_{2x+1.5y}$ stoichiometry) and additional oxygen vacancies, making the film oxygen-deficient. Our experiments show that all levels shift to higher binding energies in the presence of {\em additional} oxygen vacancies; the $E_\mathrm{B}$ values of the fully oxidized films are not influenced by the existence of Y-induced O vacancies in the film. At first glance, this may seem unexpected as both types of oxygen vacancies are positively charged V$_\mathrm{O}^{\bullet\bullet}$. The difference lies in the fact that the crystal remains electrically neutral when there are only doping-induced O vacancies, while any additional vacancies lead to an excess of positive charge, and, hence, a positive electrostatic potential of the film. As discussed previously, \cite{lacknerPCCP} this is the reason for the increased binding energy of all species in oxygen-deficient films. We have observed that the shift of the O 1s lines upon oxidation of the film is slightly less (by 0.2--0.3 eV) than the shift of the Zr and Y 3d lines. A similar behavior occurs when oxidation goes hand in hand with the tetragonal $\rightarrow$ monoclinic phase transformation and was explained with a changing band gap or the change of the local structure. \cite{lacknerPCCP} Compared to the electrostatic  core level shifts of more than 1\,eV, surface core level shifts of Zr in ZrO$_2$ are predicted to be much smaller ($\approx 0.15$\,eV, Ref.\ \onlinecite{li_growth_2015}), and we consider it likely that the same is true for Y in yttria-doped ZrO$_2$.

\section{\label{sec:concl}Conclusions}

We have studied yttria doping by Y deposition on high-purity, 5 ML-thick zirconia films. Annealing at 750\,$^\circ$C in $5 \times 10^{-7}$\,mbar \ce{O2} leads to an equilibrium distribution of the Y. In this state, the surface is strongly Y-enriched; the XPS results are not compatible with pure \ce{ZrO2} in the surface layer and Y in the second layer only as suggested previously. Doping levels of 0.5\,mol\% \ce{Y2O3} (averaged over the film thickness, higher Y content at the surface and less in the layers below) are sufficient to stabilize the tetragonal phase even for an oxidized film. We could also transform monoclinic \ce{ZrO2} films to the tetragonal phase via Y-doping (2\,mol\% \ce{Y2O3}). The films showed similar XPS core level shifts as pure \ce{ZrO2} films, with high binding energies caused by the positive electrostatic potential when the films carry net positive charge due to a concentration of positive oxygen vacancies (V$_\mathrm{O}^{\bullet\bullet}$) exceeding the one introduced by the doping. Thin yttria-doped zirconia films provide a good basis for further studies of important properties of this material, such as phase stability under more extreme conditions, reactivity, or diffusion of dopants as well as impurities, which can be added intentionally in a controlled fashion.

\section{\label{sec:ackn}Acknowledgements}

This work was supported by the Austrian Science Fund (FWF) under project number F4505 (Functional Oxide Surfaces and Interfaces -- FOXSI).

\bibliography{YSZ_bibliography}

\end{document}